\documentstyle[11pt,pasp,twoside]{article}
\def\edcomment#1{\iffalse\marginpar{\raggedright\sl#1\/}\else\relax\fi}
\reversemarginpar

\begin{document}
\title{High Resolution X-Ray Spectroscopy of Seyfert 2 Galaxies}
\author{S.~M. Kahn, A. Kinkhabwala, M. Sako, E. Behar, F.~B.~S. Paerels} 
\affil{Columbia Astrophysics Laboratory, Columbia University,
550 W. 120th Street, New York, NY  10027, USA}
\author{A.~C. Brinkman, J.~S. Kaastra, R. van der Meer}
\affil{SRON, Sorbonnelaan 2, Utrecht, 3548 CA, The Netherlands}
\author{D.~A. Liedahl}
\affil{Physics and Advanced Technology Division, Lawrence Livermore 
National Laboratory, P.O. Box 808, Livermore, CA  94550, USA}

\begin{abstract}
X-ray spectroscopy of Seyfert 2 galaxies provides an
excellent probe of the circumnuclear environment in active galactic nuclei.
The grating experiments on
both Chandra and XMM-Newton have now provided the first high
resolution spectra of several of the brightest Seyfert 2's.  We present
Chandra HETG data on Markarian 3, and XMM-Newton RGS data
on NGC 1068.  In both cases, the spectra are dominated by emission lines
due to radiative recombination following photoionization, photoexcitation,
and fluorescence.  There is no evidence for any significant contribution
from collisionally-heated gas.    
\end{abstract}

\section{Introduction}

In standard models of active galactic nuclei, Seyfert 2
galaxies are interpreted as systems viewed at sufficiently high 
inclination angles that the central continuum source is obscured by the 
surrounding dusty torus (e.~g., Krolik 1999).  However, residual soft X-ray 
emission 
is observed from many of these systems.  This must originate in
the unobscured nuclear vicinity, or perhaps in other regions of the galaxy.
While one expects some soft X-ray emission to
be produced by reprocessing of the strong continuum flux in the circumnuclear
environment, collisionally-heated gas associated
with starburst regions might also provide an important contribution 
(e.~g., Wilson et~al. 1992).

High resolution X-ray spectroscopy provides the least ambiguous
means of distinguishing between these two interpretations.  Discrete
spectral features formed in collisional plasmas are qualitatively different
from those formed in photoionized gas (Liedahl 1999).  In 
collisional plasmas, emission lines are primarily produced by electron impact 
excitation from the ground state, which favors strong resonance transitions.  
Fe L-shell transitions are especially prominent.  In photoionized plasmas, 
the electron temperature and density are too low for
excitation by electron impact, and the lines are produced 
mostly via radiative cascades following recombination, which favors the 
lowest-lying multiplet levels, thereby producing forbidden transitions.  Also, 
Fe L 
lines are much weaker than hydrogenic and heliumlike K-shell transitions of 
lower Z elements.  Another characteristic
difference involves the radiative recombination continua (RRC).  
In collisional plasmas, these are broad weak features, barely 
discernible above the bremsstrahlung continuum.  In photoionized plasmas, 
the RRC are much narrower and far more prominent (Liedahl and Paerels 1996).

The grating experiments on Chandra and XMM-Newton have provided the first
X-ray spectra of Seyfert 2's with sufficient resolution and sensitivity
to investigate these issues in detail.  Here we present Chandra HETG
spectra of Markarian 3 and XMM-Newton RGS spectra of NGC 1068.  The data 
are consistent with all emission arising in photoionized gas.
We can use these spectra to derive sensitive constraints on the structure
of the circumsource environment.

\section{Markarian 3}

The Chandra HETG spectrum of Markarian 3 is shown in Figure 1 (Sako et~al. 
2000).  A variety of narrow features are readily discernible
and are explicitly labeled in the figure.  These include Ly series 
transitions of hydrogenic O, Ne, Mg, Si, S, and Fe,
heliumlike complexes of O, Ne, Mg, Si, S, Ar,
and Fe, near-neutral fluorescence K$\alpha$ lines of Mg, Si, S, and
Fe, and a few weak Fe L lines for a range of charge states (Fe XVII through
XXIV).

\begin{figure}
\plotfiddle{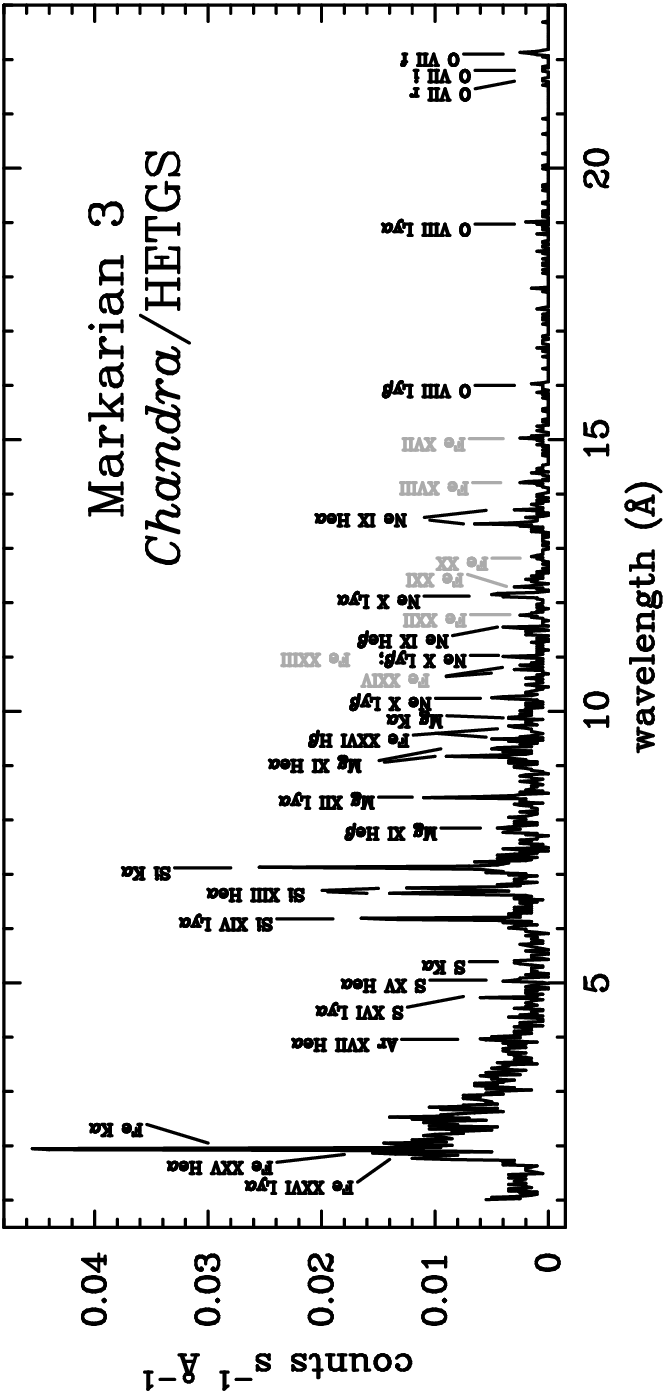}{2.2in}{-90}{60}{60}{-225}{212}
\caption{Chandra HETG spectrum of Markarian 3.}
\end{figure}

The He$\alpha$ complexes are especially interesting.  For O VII, the 
forbidden line is clearly the strongest, and the relative line ratios 
agree with the values expected for pure recombination
(Porquet \& Dubau 2000).  For both Ne IX and Si XIII, the resonance lines
are roughly comparable to the forbidden lines, but the ratios are still
not as high as expected for electron impact excitation.  This suggests
possible contributions from both recombination and collisional excitation
emission from the same plasma or spatially-distinct plasmas.  The presence 
of the Fe L features might also bolster that interpretation.

However, we are firmly convinced that this is {\em not} what is happening in
this spectrum.  Both the heliumlike resonance lines and the Fe L lines are
not produced collisionally, but instead are photoexcited by the central
continuum radiation.  Two major clues lead us to this interpretation:
(1) The Fe L lines are too weak relative to the heliumlike resonance
lines to be consistent with electron impact excitation, unless we significantly
reduce the iron abundance in the putative collisional component.  (2) The
specific Fe L lines observe are all 3d--2p and 3p--2s.  These 
have high oscillator strengths, and are the most
prominent features expected for photoexcitation.  There are no accompanying
3s-- 2p transitions, which have much lower oscillator strengths.  In 
collisional plasmas, the two sets of transitions have near comparable line
intensities.

At low to intermediate line optical depths, photoexcitation and 
recombination provide comparable contributions to  the 
X-ray line spectrum of photoionized gas.  This interpretation 
suffices for Markarian 3.  We see no conclusive evidence for a contribution
from collisional gas.

\section{NGC 1068}

The XMM-Newton RGS spectrum of NGC 1068 is shown in Figure 2 (Kinkhabwala
et~al. 2001a).  We see a very line-rich spectrum dominated by hydrogenic and 
heliumlike lines (C through Si), accompanied by weak Fe L-shell
transitions.  In addition, we see very prominent RRC.  All RRC for the 
hydrogenic and heliumlike charge states of C, N, O, and Ne are detected.

\begin{figure}
\plotfiddle{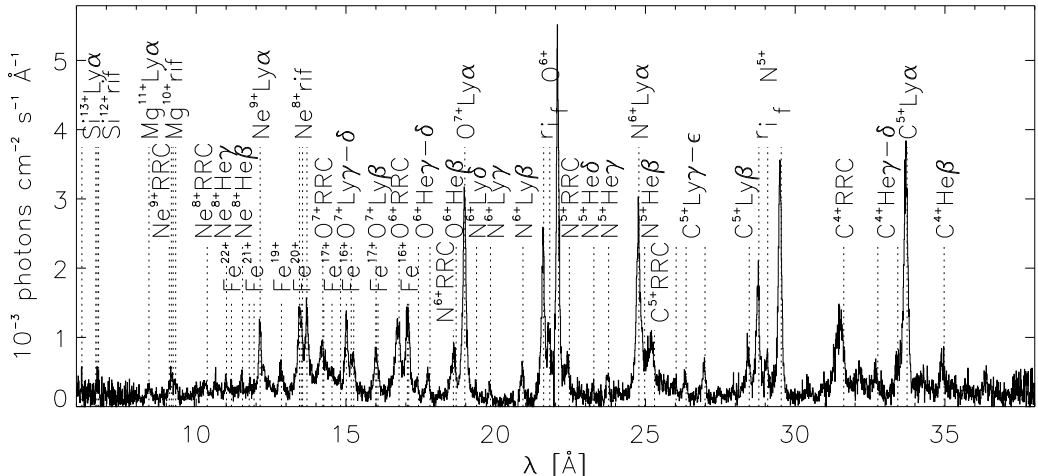}{2.1in}{90}{60}{60}{230}{-93}
\caption{XMM-Newton RGS Spectrum of NGC 1068.}
\end{figure}

The narrow RRC provide unambiguous evidence that most of the X-ray
emission we are observing is produced in photoionized gas.  This conclusion is
supported by the He$\alpha$ complexes, in which
the forbidden line is brighter than the resonance.  The
ratios of the forbidden line intensities to the RRC intensities agree with
expectations for recombination cascades.

Nevertheless, as for Markarian 3, recombination alone does not provide
a good quantitative fit to the entire NGC 1068 spectrum.  The heliumlike
resonance lines are still too bright relative to the forbidden lines, and
the Fe L lines are brighter than would be expected for purely recombining
gas.  Here again, we can explain all of these features by allowing for
additional contributions from photoexcitation.  Conclusive evidence
for this comes from the hydrogenic line series:
The higher series lines are observed to be significantly brighter relative
to Ly$\alpha$ than is expected for either recombining gas or 
collisionally-excited gas.  This is a natural consequence of photoexcitation 
at intermediate optical depths due to saturation of the highest oscillator 
strength lines.

We have developed a fully self-consistent model for the NGC 1068
spectrum in terms of a photoionized medium which correctly accounts for
both the photoexcitation and radiative recombination contributions
(Kinkhabwala et~al. 2001a, Behar et~al. 2001).  Individual column densities 
and velocity distributions for each ion are left as free
parameters.  The model provides an amazingly good fit to these high statistics
data.  Fits to the earlier Chandra HETG spectra of Markarian 3, NGC 4151 
(Ogle et~al. 2000), and the Circinus galaxy (Sambruna et~al. 2000) as well as 
to a recent observation of NGC 4507 yield similar results (Kinkhabwala et~al. 
2001b).  In particular, we find that the Ogle et~al. (2000) conclusion that 
the NGC 4151 spectrum indicates the presence of hot collisional gas is 
incorrect.  All of the features in the spectra of all Seyfert 2 galaxies so 
far observed with Chandra and XMM are consistent with fluorescence of cold gas 
and photoexcitation and recombination cascades in photoionized gas.

\section{Conclusion}

The high resolution X-ray spectra of Seyfert 2
galaxies display a wealth of emission features produced by fluorescence,
photoexcitation, and recombination cascades in radiatively-driven gas.  
There is no evidence for measureable contributions from hot 
collisionally-ionized gas in any of the sources which have been studied to
date.  Further observations of this kind should be
very helpful in elucidating the nature of the circumsource medium in active
galactic nuclei.


\begin{references}

\reference
Behar, E., et~al. 2001, in ASP Conf. Ser. Vol. TBD, Mass Outflow in Active
Galactic Nuclei: New Perspectives, ed. Crenshaw, D.M., Kraemer, S.B. \& 
George, I.M. (San Francisco: ASP), TBD

\reference
Kinkhabwala, A., et~al. 2001(a,b), in preparation

\reference
Krolik, J. 1999, Active Galactic Nuclei, Princeton University Press

\reference
Liedahl, D.~A. 1999, in X-ray Spectroscopy in Astrophysics, ed. van Paradijs, 
J., \& Bleeker, J.~A.~M. (Springer), 189

\reference
Liedahl, D.~A., \& Paerels, F. 1996, ApJ, 468, L33

\reference
Ogle, P.~M., et~al. 2000, ApJ, 545(2), L81

\reference
Porquet, D., \& Dubau, J. 2000, A\&AS, 143, 495

\reference
Sako, M., Kahn, S.~M., Paerels, F., \& Liedahl, D.~A. 2000, ApJ, 543, L115

\reference
Sambruna, R.~M., et~al. 2001, ApJ, 546(1), L13

\reference
Wilson, A.~S., Elvis, M., Lawrence, A., \& Bland-Hawthorn, J. 1992,
  ApJ, 391, L75

\end{references}
\end{document}